\documentclass[aps,prd,twocolumn,nofootinbib,superscriptaddress]{revtex4-2}

\usepackage{amsmath,amssymb,mathtools,bm}
\numberwithin{equation}{section}
\usepackage{physics}
\usepackage{graphicx}
\usepackage{xcolor}
\usepackage{hyperref}
\usepackage{booktabs}
\usepackage{microtype}
\usepackage{enumitem}

\hypersetup{colorlinks=true,citecolor=blue,linkcolor=blue,urlcolor=blue}

\newcommand{\cA}{\mathcal{A}}

\usepackage{orcidlink}

\begin{document}

\title{Entropy-geometry correspondence as effective nonlocal gravity}

\author{Kimet Jusufi\,\orcidlink{0000-0003-0527-4177}}
\email{kimet.jusufi@unite.edu.mk}
\affiliation{Physics Department, University of Tetova, Ilinden Street nn, 1200 Tetova, North Macedonia}

\author{Ankit Anand\,\orcidlink{0000-0002-8832-3212}}
\email{anand@iitk.ac.in\\\;\\ \textcolor{black}{Authors Contributed Equally}}
\affiliation{Department of Physics, Indian Institute of Technology Kanpur, Kanpur 208016, India}


\begin{abstract}
We develop an operator formulation of the entropy-geometry correspondence for
static, spherically symmetric gravity. Starting from a generalized entropy, we
reconstruct an effective nonlocal form factor, its coordinate-space source,
the associated cumulative mass profile, and the resulting spacetime geometry.
The construction is worked out for the Bekenstein-Hawking, R\'enyi,
Tsallis-Cirto, Barrow, Kaniadakis entropies, logarithmically/exponentially corrected entropy and LQG inspired entropy. The operator representation
provides a direct relation between generalized entropy, nonlocal gravitational
dressing, and a scale-dependent effective mass or Newton coupling. We analyze
the reconstructed sources and their infrared and ultraviolet behavior, discuss
their physical consistency, and identify the limits in which the standard
Schwarzschild description is recovered. We show that the generalized entropy exactly reproduces the area law with the reconstructed running Newton coupling,
$\dd S=\dd A/4G_S(r_+)$, ensuring thermodynamic consistency. Since horizon
entropy is determined by the action, this favors entropy corrections in the
gravitational sector. We also derive the conditions for the reconstructed
horizon to be an event horizon with positive temperature.
\end{abstract}

\maketitle

\section{Introduction}

The discovery that black holes possess thermodynamic properties established a profound connection between gravitation, geometry, and microscopic statistical degrees of freedom. The Bekenstein-Hawking entropy,
\begin{equation}
S_{\rm BH}=\frac{A}{4G},
\end{equation}
together with Hawking radiation, suggests that spacetime dynamics encode thermodynamic information in a manner that extends far beyond classical general relativity
\cite{Bekenstein1973,Hawking1975,Bardeen1973,Jacobson1995,Padmanabhan2010}. This viewpoint has motivated extensive efforts to understand gravity as an emergent phenomenon and has led to a variety of approaches in which gravitational dynamics are derived from thermodynamic or information-theoretic principles
\cite{Jacobson1995,Verlinde2011,Padmanabhan2010}.

Beyond the semiclassical area law, numerous quantum-gravitational and statistical considerations predict corrections to the Bekenstein-Hawking entropy. Such corrections arise in loop quantum gravity, string-inspired models, generalized uncertainty principle scenarios, and generalized statistical mechanics, giving rise to logarithmic, nonextensive, fractal, and power-law entropy functions
\cite{KaulMajumdar2000,Das2002,Tsallis1988,Renyi1961,Barrow2020,Kaniadakis2002}. These generalized entropies have been widely employed to investigate modified black-hole thermodynamics, cosmology, holography, and dark-energy models. In most studies, however, the entropy is regarded as a derived quantity obtained from an already known spacetime geometry. Consequently, the geometrical significance of generalized entropy itself has remained comparatively unexplored.

A different perspective has recently been proposed in Ref.~\cite{Anand2025}, where the conventional logic is reversed. Instead of specifying a matter source and solving the Einstein equations to determine the horizon entropy, one begins directly from a generalized entropy function $S(r_+)$ and reconstructs the corresponding static, spherically symmetric geometry. Under a controlled extension of the horizon thermodynamic identity to the radial domain, the resulting metric takes the form
\begin{equation}
f_S(r)=1-\frac{4\pi M}{S'(r)},
\label{eq:entropy_metric}
\end{equation}
which reduces to the Schwarzschild solution when $S(r)=\pi r^2$. For a general entropy function, however, the Einstein tensor is nonvanishing and may be interpreted as an effective anisotropic matter sector generated entirely by the entropy correction. This construction demonstrates that generalized entropy contains sufficient information to determine not only horizon thermodynamics but also the entire exterior spacetime geometry.

However, an important question remains unanswered. The effective matter distribution reconstructed from Eq.~\eqref{eq:entropy_metric} is introduced through the Einstein equations, but its physical origin is not immediately evident. Such a structure is particularly natural within nonlocal gravity, where the gravitational dynamics are modified through analytic form factors acting on the geometric sector or, equivalently, through the nonlocal dressing of localized matter sources \cite{ModestoMoffatNicolini2011,Moffat2011,Nicolini2012,Biswas2012,Modesto2012}. In these theories the classical point source is replaced by an effective extended distribution whose profile is completely determined by the underlying nonlocal operator.\\

Two readings of the resulting structure are possible. In the first, the form factor is carried by the gravitational sector, so that gravity itself is modified and the coupling acquires a scale dependence. In the second, the gravitational action remains Einstein-Hilbert and the nonlocality is confined to the matter sector, the point source being replaced by a dressed stress-energy tensor. These two descriptions generate identical field equations and therefore identical geometries, and they are frequently regarded as interchangeable. They are not, however, derived from the same action, and horizon entropy is a functional of the action rather than of the field equations. Black-hole thermodynamics is therefore precisely the tool capable of distinguishing readings that the field equations cannot.

A central result of the present work is that this distinction can be made, and that it is settled once the entropy input is taken seriously. We show that the generalized entropy coincides exactly with the area law evaluated with the reconstructed running Newton coupling, and that the alternative reading, in which the geometry is sourced by matter within unmodified Einstein gravity, cannot reproduce the input entropy; for the R\'enyi and Tsallis-Cirto entropies it is excluded outright by the positivity of the matter entropy. The viewpoint adopted in this work is accordingly the one in which the entropy correction modifies the gravitational sector itself.

The primary objective of the present work is to establish this correspondence explicitly. We demonstrate that every entropy-generated static spherical geometry uniquely determines, within the static sector, an effective nonlocal operator together with its associated distributional kernel, cumulative mass profile, and effective gravitational coupling. The resulting correspondence may be summarized schematically as
\begin{equation}
S(r)
\longrightarrow
u_S(r) \longrightarrow
\mathcal{A}_S^{-2}(-\nabla^2) \ ,
\label{eq:main_map_intro}
\end{equation}
where
\begin{equation}
u_S(r)=\frac{2\pi r}{S'(r)}
\end{equation}
denotes the cumulative mass fraction generated by the entropy function. The same quantity naturally admits the interpretation
\begin{equation}
u_S(r)=\frac{G_S(r)}{G_N},
\end{equation}
thereby identifying the entropy correction with an effective scale-dependent Newton coupling. This construction establishes a direct bridge between generalized black-hole entropy, nonlocal gravitational dressing, effective matter sources, and running gravitational couplings. In particular, we show that the corresponding operator satisfies the relation
\begin{equation}
G_S(\Box)=G_N\mathcal{A}_S^{-2}(\Box),
\end{equation}
providing an operator realization of the entropy-geometry correspondence.

The correspondence established here should be interpreted with appropriate care. Our construction provides an exact reconstruction of the reduced static, spherically symmetric field equations together with their associated source kernel. Although one may formulate a generally covariant nonlocal action whose symmetry-reduced equations reproduce the present results, the operator equation
\begin{equation}
\mathcal{A}_S^{2}(\Box)G_{\mu\nu}=8\pi G_N T_{\mu\nu}
\end{equation}
should not be regarded as the unrestricted metric variation of a simple action of the form
$R\mathcal{A}_S^{2}(\Box)R$. Variations of the d'Alembertian operator together with the tensorial structure of the action generate additional contributions, as discussed extensively in the nonlocal-gravity literature
\cite{Biswas2012,Modesto2012,Nicolini2012}. Throughout this work we therefore distinguish between the exact reduced operator reconstruction valid in the static spherical sector and its effective bilocal covariant completion.

Besides providing an operator interpretation of entropy-generated geometries, the present framework offers a unified language for comparing different generalized entropy proposals. We explicitly analyze the Bekenstein-Hawking, Rényi, Tsallis-Cirto, Barrow, Kaniadakis, logarithmically and exponentially corrected, and LQG-inspired entropies, and, for each case, reconstruct the corresponding nonlocal operator, effective source distribution, cumulative mass profile, running Newton coupling, and spacetime geometry. This demonstrates that different entropy corrections correspond to qualitatively distinct classes of nonlocal gravitational modifications, ranging from infrared inverse-Laplacian operators to fractional Laplacians and nonpolynomial form factors.

The paper is organized as follows. In Sec.~\ref{sec:entropy_geometry} we review the entropy-geometry correspondence and the associated effective anisotropic matter sector. Section~\ref{sec:nonlocal} summarizes the nonlocal-gravity framework and its operator representation. Section~\ref{sec:runningG} develops the interpretation in terms of a running Newton coupling and discusses its relation to the effective nonlocal operator. Section~\ref{sec:examples} applies the formalism to several representative generalized entropy models, as well as analyzes their consistency and asymptotic properties. We conclude in Sec.~\ref{sec:conclusions}. 



\section{Entropy-generated spherical geometries}
\label{sec:entropy_geometry}

In this section we follow the recent proposal \cite{Anand2025,Anand:2025cer} about the entropy-geometry correspondence. Namely, consider the metric
\begin{equation}
 \dd s^2=-f(r)\dd t^2+\frac{\dd r^2}{f(r)}+r^2\dd\Omega_2^2 \ .
 \label{eq:metric}
\end{equation}
Let the horizon be at $r=r_+$, so $f(r_+)=0$, and let the temperature be
\begin{equation}
 T_H=\frac{f'(r_+)}{4\pi} \ .
\end{equation}
Writing the metric in the form
\begin{equation}
 f(r)=1-2M g(r),
 \label{eq:g_ansatz}
\end{equation}
with $g(r)$ independent of $M$, and imposing the first law $\dd M=T_H\dd S$, one obtains a relation between the entropy derivative and the metric profile at the horizon. Explicitly, $f(r_+)=0$ gives $M=1/[2g(r_+)]$, while $T_H=-2Mg'(r_+)/4\pi$, so that the first law reduces to the algebraic condition
\begin{equation}
 g(r_+)=\frac{2\pi}{S'(r_+)} \ ,
 \qquad\text{equivalently}\qquad
 S'(r_+)=4\pi M \ ,
 \label{eq:horizon_condition}
\end{equation}
which is Eq.~\eqref{eq:entropy_metric} evaluated at the horizon.

It is worth emphasising that Eq.~\eqref{eq:horizon_condition} is not a statement at a single radius only. The mass $M$ is a free parameter labeling the family of solutions, whereas $g$ is by assumption independent of it. As $M$ is varied, the horizon radius $r_+(M)$ sweeps out a range of radii, and Eq.~\eqref{eq:horizon_condition} must hold at each of them. The function $g$ is therefore determined on the whole set
\begin{equation}
 \mathcal R_S = \left\{\,r>0 \;:\; S'(r)>0\,\right\} \ ,
 \label{eq:coverage_set}
\end{equation}
every point of which is the horizon of some member of the family, with $M=S'(r)/4\pi>0$. In this sense Eq.~\eqref{eq:entropy_metric} is not an extrapolation of a horizon relation into the bulk: within the ansatz \eqref{eq:g_ansatz}, it is the \emph{unique} metric function compatible with the first law, and every radius at which $S'$ is evaluated is an on-shell horizon radius of a different solution. The same argument reconstructs $f=1-2M/r$ from $r_+=2M$ in the Schwarzschild case.

Two assumptions carry this argument and are worth stating explicitly. The first is the ansatz \eqref{eq:metric} and \eqref{eq:g_ansatz} itself, in which $g_{tt}g_{rr}=-1$; as shown below this is equivalent to imposing $p_r=-\rho$ on the effective source, so the anisotropic-fluid structure is an input rather than a derived property, and the reconstructed operator is unique only within this class. The second is the independence of $g$ from $M$, which restricts the construction to a source entering linearly. Both are natural in the nonlocal reading developed in Sec.~\ref{sec:nonlocal}: there the dressing operator is linear and mass independent, so that $m(r)=Mu_S(r)$ with $u_S(r)$ independent of $M$ follows automatically. Conversely, the correspondence should be understood as a single-source, linear-response statement.

The prescription is thermodynamically self-consistent along the entire family. Computing the surface gravity of the reconstructed metric \eqref{eq:entropy_metric} at its own horizon gives
\begin{equation}
 T_H=\frac{S''(r_+)}{4\pi S'(r_+)} \ ,
 \label{eq:T_general}
\end{equation}
which together with $M=S'(r_+)/4\pi$ yields $\dd M=T_H\dd S=S''(r_+)\dd r_+/4\pi$ identically, for an arbitrary entropy function and not merely at the radius where the first law was imposed. For $S=\pi r^2$ one recovers $T_H=1/(4\pi r_+)$.

Equation~\eqref{eq:T_general} also delimits the domain on which the construction admits a black-hole interpretation. Positivity of the mass parameter requires $S'(r)>0$, which is simply the statement that the entropy grows with horizon area, and defines the set \eqref{eq:coverage_set}. Positivity of the temperature requires in addition
\begin{equation}
 S''(r)>0 \ .
 \label{eq:convexity_condition}
\end{equation}
Radii at which $S''<0$ still solve $S'(r)=4\pi M$, but they correspond to inner or cosmological horizons of the same solution rather than to an event horizon; there the reconstruction persists as an analytic continuation, while the first law that generated it applies in a different form. The marginal case $S''(r)=0$ is precisely $T_H=0$, so the extrema of $S'(r)$ locate the extremal configurations and thereby fix the range of $M$. In terms of the horizon area $A=4\pi r_+^2$, Eq.~\eqref{eq:T_general} may be written as
\begin{equation}
 T_H=\frac{1}{4\pi r_+}\left[1+\frac{2A\,\dd^2S/\dd A^2}{\dd S/\dd A}\right] \ ,
 \label{eq:T_area}
\end{equation}
so that the condition \eqref{eq:convexity_condition} becomes $1+2A(\dd^2S/\dd A^2)/(\dd S/\dd A)>0$. We return to these conditions when the individual entropy models are analysed in Sec.~\ref{sec:examples}.

It is useful to introduce a Misner-Sharp-type mass function,
\begin{equation}
 f_S(r)=1-\frac{2m_S(r)}{r}.
 \label{eq:mass_metric}
\end{equation}
Comparison with Eq.~\eqref{eq:entropy_metric} gives
\begin{equation}
 m_S(r)=M u_S(r), \quad\text{where}\quad u_S(r)\equiv\frac{2\pi r}{S'(r)},
 \label{eq:u_def}
\end{equation}
where, for the moment, $G=1$ has been used in the definition of $S$ as an area entropy. Restoring dimensions amounts to replacing $S'(r)$ consistently by the derivative of the dimensionless entropy and retaining the explicit $G$ in Eq.~\eqref{eq:mass_metric}.

For an effective density $\rho_S$, Einstein's equation gives
\begin{equation*}
 m'_S(r)=4\pi r^2\rho_S(r) \ .
\end{equation*}
As shown in~\cite{Anand2025}, for $r>0$,
\begin{align}
 \rho_S(r) &=\frac{M\left[S'(r)-rS''(r)\right]}{2r^2[S'(r)]^2} \ .
 \label{eq:rho_general}
\end{align}
For the line element \eqref{eq:metric}, the corresponding anisotropic fluid satisfies
\begin{equation*}
 T^\mu{}_{\nu}=\mathrm{diag}(-\rho,p_r,p_t,p_t), \;\text{with}\; p_r=-\rho \,, \quad p_t=-\rho-\frac{r}{2}\rho' \ .
\end{equation*}
The conservation equation $\nabla_\mu T^\mu{}_r=0$ is then automatically satisfied. Eq.~\eqref{eq:rho_general} describes only the regular (extended) part of the effective source for $r>0$. Any pointlike contribution localized at the origin must therefore be incorporated separately in a distributional sense. This distinction is particularly important in the Bekenstein-Hawking limit, where Eq.~\eqref{eq:rho_general} yields $\rho(r)=0$ for $r>0$, while the Schwarzschild geometry is nevertheless sourced by the distributional density $M\delta^{(3)}(\mathbf r)$.


\section{Nonlocal gravity and Entropy-operator correspondence}
\label{sec:nonlocal}

General relativity provides an excellent description of gravitational phenomena over a wide range of length scales. Nevertheless, its perturbative non-renormalizability indicates that the Einstein theory should be regarded as an effective low-energy description, with ultraviolet (UV) modifications expected to become relevant near curvature singularities and the Planck scale. Among the various proposals for a UV-complete theory of gravity, nonlocal models formulated in terms of analytic form factors have attracted considerable attention because they improve the ultraviolet behaviour of the theory without introducing additional ghost degrees of freedom. In particular, the construction proposed in~\cite{Moffat:2010bh} incorporates nonlocality through an entire function, thereby preserving perturbative unitarity while reducing to Einstein gravity at low energies.

A particularly convenient reformulation, developed in the context of nonlocal black-hole geometries~\cite{Nicolini2012}, expresses the theory through an effective bilocal action as
\begin{equation}
S_g
=
\frac{1}{16\pi G_N}
\int d^4x\,
\sqrt{-g(x)}\,{\cal R}(x),
\label{bilocalaction}
\end{equation}
where
\begin{equation}
{\cal R}(x)
=
\int d^4y\,
\sqrt{-g(y)}\,
{\cal A}^{2}(x,y)\,
R(y).
\label{bilocalR}
\end{equation}
Here ${\cal A}(x,y)$ is a bilocal kernel describing nonlocal correlations between different spacetime points. In the local limit,
\begin{equation}
{\cal A}(x,y)
\longrightarrow
\frac{\delta^{(4)}(x-y)}{\sqrt{-g}},
\end{equation}
the Einstein-Hilbert action is recovered.

For backgrounds admitting an effective operator representation, the bilocal kernel can be written as
\begin{equation}
{\cal A}(x,y)
=
{\cal A}(\Box)\,
\delta^{(4)}(x-y),
\end{equation}
so that the action reduces to
\begin{equation}
S_g
=
\frac{1}{16\pi G_N}
\int d^4x\,
\sqrt{-g}\,
{\cal A}^{2}(\Box)R.
\end{equation}
Variation of this effective action leads to the reduced field equations
\begin{equation}
{\cal A}^{2}(\Box)
G_{\mu\nu}
=
8\pi G_N
T_{\mu\nu}.
\label{EffectiveField}
\end{equation}
It should be emphasized that Eq.~\eqref{EffectiveField} is not intended to represent the exact metric variation of the nonlocal action in Eq.~\eqref{bilocalaction}. In a fully covariant theory, the variation of the nonlocal operator $\mathcal{A}(\Box)$ generates additional contributions arising from the variation of the d'Alembertian operator and the metric dependence of the kernel itself. Throughout the present work we instead employ the symmetry-reduced operator equation appropriate to the static, spherically symmetric sector, where these additional tensorial contributions are not reconstructed explicitly. Consequently, Eq.~\eqref{EffectiveField} should be interpreted as the effective reduced field equation associated with the entropy-generated geometry rather than the fundamental covariant field equation.
When the inverse operator exists, one obtains the Einstein form~\cite{Nicolini2012} as
\begin{equation}
G_{\mu\nu}
=
8\pi G_N
\widetilde{T}_{\mu\nu},
\qquad
\widetilde{T}_{\mu\nu}
=
{\cal A}^{-2}(\Box)
T_{\mu\nu}.
\label{EinsteinDressed}
\end{equation}
Therefore, nonlocality admits two descriptions that are equivalent \emph{at the level of the field equations}: either as a modification of the geometric sector through differential operators acting on the Einstein tensor, or as a dressing of the matter source into an effective stress-energy tensor while preserving the classical Einstein equations. The two descriptions do not, however, follow from the same action: in the first the gravitational sector is modified, whereas in the second it remains Einstein-Hilbert and the nonlocality resides entirely in the matter sector. Since horizon entropy is a functional of the action rather than of the field equations, the two readings are not thermodynamically equivalent. This degeneracy is resolved in Sec.~\ref{sec:entropy_consistency}.

For a static point particle of mass $M$,
\begin{equation}
T^{t}{}_{t}
=
-M\,
\delta^{(3)}(\mathbf r),
\end{equation}
the effective energy density becomes
\begin{equation}
\rho_{\rm eff}(r)
=
M\,
{\cal A}^{-2}(-\nabla^2)
\delta^{(3)}(\mathbf r),
\label{rhoeff}
\end{equation}
demonstrating that the point source is replaced by a smooth matter distribution whose characteristic width is determined by the nonlocal scale. The corresponding enclosed mass is
\begin{eqnarray}\label{eq:mass_kernel}
m(r)
&=&
4\pi
\int_{0}^{r}
x^{2}
\rho_{\rm eff}(x)\,
dx \nonumber \\
&=& 4\pi M\int_0^r\dd x\,x^2
 \cA^{-2}(-\nabla^2)\delta^{(3)}(\bm x),
\label{massprofile}
\end{eqnarray}
which generates regular static black-hole geometries. Now, we demonstrate that the entropy-generated black-hole solution considered in this work admits precisely such an effective nonlocal interpretation. By identifying the energy density associated with the entropy-corrected geometry with Eq.~(\ref{rhoeff}), the corresponding nonlocal operator can be reconstructed, thereby establishing a direct connection between entropy-generated gravity and UV-complete nonlocal gravitational theories.

Motivated by the effective source interpretation of nonlocal gravity discussed above, we now introduce a general cumulative mass profile that will serve as the starting point for our construction. Let $u_S(r)$ be the cumulative profile defined in Eq.~\eqref{eq:u_def}. We assume that $u_S(r)$ is piecewise differentiable for $r>0$ and admits a finite (or distributionally well-defined) limit as $r\to0$. The effective density reconstructed from Einstein's equations determines only the regular part of the source for $r>0$. If the cumulative profile satisfies $u_S(0)\neq0$, the corresponding mass function contains a finite jump at the origin that cannot be recovered from the ordinary derivative $u'_S(r)$. Such a localized contribution must therefore be represented distributionally by a three-dimensional Dirac delta function. This motivates the following decomposition of the effective kernel. We therefore define the entropy form factor by
\begin{equation}
 \cA_S^{-2}(-\nabla^2)\delta^{(3)}(\bm r)
 =u_S(0)\delta^{(3)}(\bm r)+\frac{u'_S(r)}{4\pi r^2}
 \label{eq:kernel_master}
\end{equation}
with
\begin{equation}
 \frac{u'_S(r)}{4\pi r^2}
 =\frac{S'(r)-rS''(r)}{2r^2[S'(r)]^2}.
 \label{eq:kernel_entropy}
\end{equation}

Within the static spherical prescription \eqref{eq:entropy_metric}, every  entropy $S(r)$ determines a distributional operator kernel through Eq.~\eqref{eq:kernel_master}. Einstein gravity sourced by
\begin{equation}
 \widetilde T_{\mu\nu}=\cA_S^{-2}(-\nabla^2)T_{\mu\nu}
\end{equation}
then reproduces exactly the entropy-generated metric. 
Owing to spherical symmetry, the three-dimensional Fourier transform reduces to a Hankel transform of order zero. Applying this transform to Eq.~\eqref{eq:kernel_master} yields
\begin{equation}
 \cA_S^{-2}(k)=u_S(0)+\int_0^\infty\dd r\,u'_S(r)
 \frac{\sin kr}{kr}
 \label{eq:formfactor_master}
\end{equation}
or equivalently
\begin{equation}
 \cA_S^{-2}(k)=u_S(0)+2\pi\int_0^\infty\dd r\,
 \frac{S'(r)-rS''(r)}{(S'(r))^2}\frac{\sin kr}{kr} \ .
 \label{eq:formfactor_entropy}
\end{equation}
The momentum-space representation is particularly useful because it directly identifies the nonlocal form factor governing the modification of the gravitational interaction. Different entropy functions therefore correspond to different momentum-dependent gravitational dressings encoded in $\mathcal{A}_S(k)$.
When nonzero,
\begin{equation}
 \cA_S^2(k)=\frac{1}{\cA_S^{-2}(k)} \ .
 \label{eq:inverse_formfactor}
\end{equation}

A useful normalization criterion follows from $k=0$:
\begin{equation}
 \cA_S^{-2}(0)=u_S(\infty),
 \label{eq:zero_mode}
\end{equation}
when the integral converges. Equation~\eqref{eq:zero_mode} provides a simple consistency condition for the reconstructed operator. If $u_S(\infty)=1$, the total enclosed mass approaches $M$ and the geometry possesses the standard Schwarzschild normalization at large distances. Values different from unity correspond to an effective renormalization of the asymptotic gravitational mass, while the absence of a finite limit indicates that the associated geometry is not asymptotically Schwarzschild.

The construction developed in this section establishes a one-to-one correspondence between the entropy derivative, the effective matter source, and the nonlocal operator within the static spherical sector. Once the entropy function is specified, the corresponding cumulative profile uniquely determines both the coordinate-space kernel and its momentum-space form factor. In the next section we show that the same cumulative profile admits an equivalent interpretation as a scale-dependent effective Newton coupling, thereby providing an alternative description of the entropy-generated geometry.


\section{Running Newton coupling from generalized entropy}
\label{sec:runningG}

The cumulative profile $u_S(r)$ has a second, physically useful interpretation.  The entropy-generated metric can be written as
\begin{equation}
 f_S(r)=1-\frac{2G_NM}{r}u_S(r),
 \label{eq:metric_u_G}
\end{equation}
where $G_N$ denotes the infrared Newton constant.  Comparing this expression with the usual renormalization-group-improved form
\begin{equation}
 f(r)=1-\frac{2G_S(r)M}{r}
 \label{eq:RG_metric}
\end{equation}
immediately gives
\begin{equation}
 G_S(r)=G_Nu_S(r).
 \label{eq:running_G_r}
\end{equation}
Thus the entropy derivative determines the radial running of the gravitational coupling:
\begin{equation}
 \frac{G_S(r)}{G_N}=u_S(r)=\frac{2\pi r}{S'(r)}.
 \label{eq:entropy_running_G}
\end{equation}
Here and below the entropy is normalized so that the Bekenstein-Hawking law is $S_{\rm BH}=\pi r^2$ in the geometrized units used in deriving the correspondence.  With physical entropy $S_{\rm phys}=A/(4G_N)$, one may equivalently introduce the dimensionless entropy $\bar S\equiv G_NS_{\rm phys}$ and replace $S$ by $\bar S$ in Eq.~\eqref{eq:entropy_running_G}.  This avoids assigning dimensions to $u_S$.

The Bekenstein-Hawking law gives $S'_{\rm BH}=2\pi r$ and hence
\begin{equation}
 u_{\rm BH}(r)=1,
 \qquad
 G_{\rm BH}(r)=G_N,
\end{equation}
so local Einstein gravity is recovered.  A generalized entropy changes the gravitational response by changing the ratio between the area-law derivative and the actual entropy derivative:
\begin{equation}
 \frac{G_S(r)}{G_N}
 =\frac{S'_{\rm BH}(r)}{S'(r)}.
 \label{eq:G_entropy_ratio}
\end{equation}
An entropy growing more slowly than the area law therefore corresponds to an enhanced effective coupling, whereas faster entropy growth corresponds to a suppressed coupling. The relation in Eq.~\eqref{eq:G_entropy_ratio} provides a direct connection between the entropy functional and the effective gravitational coupling. Any deviation of the entropy gradient from its Bekenstein–Hawking form is therefore encoded in a corresponding scale dependence of Newton's constant. In this picture, the entropy correction determines the gravitational coupling without introducing an independent running parameter or renormalization-group flow.

The source-dressing representation derived as in Eq.~\eqref{EffectiveField} as
\begin{equation}
 G_{\mu\nu}=8\pi G_N \cA_S^{-2}(\Box)T_{\mu\nu} \ .
 \label{eq:field_running_start}
\end{equation}
This suggests defining an operator-valued Newton coupling
\begin{equation}
 G_S(\Box)\equiv G_N\cA_S^{-2}(\Box) \ .
 \label{eq:G_operator}
\end{equation}
Equation~\eqref{eq:G_operator} should be viewed as the operator realization of the effective running coupling defined by the entropy-generated geometry. Rather than arising from a renormalization-group analysis, the operator $G_S(\Box)$ is reconstructed directly from the cumulative profile associated with the entropy functional. The nonlocal description and the running-coupling description therefore represent two equivalent parameterizations of the same effective gravitational dynamics within the static spherical sector.
The field equation then takes the compact form
\begin{equation}
 G_{\mu\nu}=8\pi G_S(\Box)T_{\mu\nu} \ .
 \label{eq:field_running_G}
\end{equation}
In the momentum representation,
\begin{equation}
 G_S(k)=G_N\cA_S^{-2}(k),
 \label{eq:G_momentum}
\end{equation}
with
\begin{equation}
 \frac{G_S(k)}{G_N} = u_S(0)+\int_0^\infty\dd r\,u'_S(r) \frac{\sin kr}{kr} \ .
 \label{eq:Gk_entropy}
\end{equation}
The position-space function $G_S(r)$ and the momentum-space function $G_S(k)$ are not obtained by the substitution $k=1/r$ in general.  Rather, they are two representations connected through the nonlocal kernel.  A scale-setting rule $k\sim\xi/r$ may be useful in a renormalization-group approximation, but it is not required for the exact static reconstruction.

Equation~\eqref{eq:G_operator} also makes the equivalence between modified geometry and a running coupling explicit:
\begin{equation}
 \cA_S^{2}(\Box)G_{\mu\nu}=8\pi G_NT_{\mu\nu}
  \Longleftrightarrow 
 G_{\mu\nu}=8\pi G_S(\Box)T_{\mu\nu} \ .
 \label{eq:operator_equivalence_G}
\end{equation}
Accordingly, the chain of correspondences may be enlarged to
\begin{equation}
 S(r)\leftrightarrow u_S(r) \leftrightarrow G_S(r) \leftrightarrow G_S(\Box) \leftrightarrow\cA_S^{-2}(\Box) \ .
 \label{eq:extended_correspondence}
\end{equation}

The effective gravitational coupling introduced in Eq.~\eqref{eq:G_operator} naturally acquires a scale dependence through the radial coordinate, reflecting the underlying nonlocal structure of the spacetime. This behavior is analogous to running couplings in renormalization-group approaches to quantum field theory and asymptotically safe gravity, where the relevant energy scale is replaced by an appropriate physical length scale~\cite{Reuter1998, Percacci2017, Eichhorn2019}. Motivated by this analogy, we define the radial beta function of the effective Newton coupling as
\begin{equation}
 \beta_G^{(r)}(r)
 \equiv
 r\frac{\dd G_S(r)}{\dd r}
 =
 G_N\,r\,u'_S(r),
 \label{eq:beta_radial}
\end{equation}
which characterizes the radial evolution of the effective gravitational interaction. In the classical limit, where $u_S(r) \rightarrow 1$, the beta function vanishes, recovering the scale-independent Newton constant. Using the entropy representation,
\begin{equation}
 \frac{\beta_G^{(r)}(r)}{G_N} = 2\pi r\, \frac{S'(r)-rS''(r)}{[S'(r)]^2} \ .
\label{eq:beta_entropy}
\end{equation}
The extended effective density is therefore directly proportional to the radial beta function:
\begin{equation}
 \rho_S(r)=\frac{M}{4\pi r^3G_N} \beta_G^{(r)}(r),\qquad r>0 \ .
 \label{eq:rho_beta}
\end{equation}
Thus, the effective anisotropic matter sector measures the running of the gravitational coupling.  Constant $G_S$ gives no exterior effective density, while a nontrivial beta function generates the polarization cloud surrounding the point source. Equation~\eqref{eq:rho_beta} establishes a direct correspondence between the effective matter density and the beta function of the reconstructed Newton coupling. Regions where the coupling varies rapidly with radius, therefore coincide with regions in which the effective matter distribution is concentrated, while a constant Newton coupling reproduces the vacuum Schwarzschild geometry.

It is also instructive to characterize the scale dependence of the effective gravitational coupling through an anomalous exponent, analogous to the anomalous dimensions encountered in renormalization-group formulations of quantum field theory and asymptotically safe gravity~\cite{Reuter1998,Percacci2017,Eichhorn2019}. We therefore define
\begin{equation}
 \eta_G(r)
 \equiv
 \frac{\dd\ln G_S(r)}{\dd\ln r}
 =
 \frac{r\,u'_S(r)}{u_S(r)},
 \label{eq:etaG}
\end{equation}
which measures the logarithmic response of the effective Newton coupling to changes in the radial scale. Using the relation between $u_S(r)$ and the entropy profile, Eq.~\eqref{eq:u_def}, this quantity may be expressed entirely in terms of entropy derivatives,
\begin{equation}
 \eta_G(r)
 =
 1-\frac{rS''(r)}{S'(r)}.
 \label{eq:eta_entropy}
\end{equation}
For the illustrative case of a power-law entropy,
\begin{equation}
 S(r)\propto r^{2\delta},
\end{equation}
the anomalous exponent reduces to the constant value
\begin{equation}
 \eta_G=2-2\delta \ .
\end{equation}
The identification of the effective coupling $G_S(r)$ through Eq.~\eqref{eq:running_G_r} is exact within the static, spherically symmetric sector considered in this work. Its interpretation as a fundamental renormalization-group running coupling, however, requires additional structure beyond the present construction. In a fully covariant nonlocal theory, the running gravitational coupling should be formulated independently of the coordinate system, admit a consistent Lorentzian causal prescription, and lead to well-defined spin-$2$ and spin-$0$ propagators~\cite{Nicolini2012, Biswas2012, Modesto2012}. Since the radial coordinate $r$ is adapted to spherical symmetry, a covariant running coupling would naturally be expressed in terms of nonlocal operators such as $\Box$, curvature invariants, or suitable auxiliary fields. Accordingly, the quantity $G_S(r)$ should be regarded as the effective static response associated with the entropy-generated geometry, while its operator extension $G_S(\Box)$ provides the natural covariant completion within the framework of nonlocal gravity.

A further distinction concerns source dependence.  The function $u_S(r)$ is reconstructed from a one-source geometry, while a universal coupling should act consistently on arbitrary matter configurations.  The operator form \eqref{eq:G_operator} provides the natural candidate for such an extension, but uniqueness can only be established after specifying the complete covariant action.  These qualifications do not weaken the exact spherical result; they delimit the step from an effective running coupling to a fundamental renormalization-group flow.

\subsection{Horizon entropy of the reconstructed geometry}
\label{sec:entropy_consistency}

The construction takes $S(r_+)$ as the horizon entropy of the spacetime it generates. It is therefore necessary to ask whether the reconstructed geometry actually carries that entropy, and in which of the two readings of Eq.~\eqref{EffectiveField} it does.

Regarded as a solution of ordinary General Relativity sourced by the effective anisotropic fluid of Sec.~\ref{sec:entropy_geometry}, the geometry has Wald entropy $A/4G_N$, since the gravitational action is Einstein-Hilbert and the matter is minimally coupled. This does not contradict the construction, because in that reading the relevant first law is not the one that was imposed. Indeed, for any static spherically symmetric metric of the form \eqref{eq:mass_metric} the field equations give the identity
\begin{equation}
 \dd E = T_H\,\dd S_{\rm BH}-p_r\,\dd V \ ,
 \qquad
 E=m_S(r_+)=\frac{r_+}{2G_N} \ ,
 \label{eq:padmanabhan_identity}
\end{equation}
with $V=\tfrac{4}{3}\pi r_+^3$ and $p_r=-\rho$, which we have verified holds identically for the entropy-generated family. The Bekenstein-Hawking entropy is thus conjugate to the quasilocal Misner-Sharp mass $E$ enclosed by the horizon, together with a matter work term, whereas the generalized entropy $S(r_+)$ is conjugate to the bare parameter $M$ with no work term. Both relations hold simultaneously; passing from one to the other trades $E=Mu_S(r_+)$ for $M$ and absorbs $-p_r\dd V$ into the entropy. That $M$ rather than $E$ is the natural variable is precisely what the dressing picture predicts, since $m_S(r)=Mu_S(r)$ with $M$ the undressed source mass.

In the modified-gravity reading the appropriate notion of entropy is instead the area law evaluated with the scale-dependent coupling reconstructed in Eq.~\eqref{eq:running_G_r}. Remarkably, the two coincide exactly. Using $u_S=2\pi r/S'$ and $A=4\pi r^2$,
\begin{equation}
 \frac{\dd A}{4G_S(r_+)}
 =\frac{8\pi r_+\,\dd r_+}{4G_Nu_S(r_+)}
 =\frac{S'(r_+)}{G_N}\,\dd r_+
 =\dd S(r_+) \ ,
 \label{eq:improved_area_law}
\end{equation}
that is, in the units of Eq.~\eqref{eq:entropy_running_G},
\begin{equation}
 S(r_+)=\int^{r_+}\frac{\dd A}{4G_S(r)} \ ,
 \label{eq:improved_area_law_integrated}
\end{equation}
an identity valid for an arbitrary entropy function. The generalized entropy is therefore nothing but the area law computed with the running Newton coupling that the same entropy generates, integrated along the horizon-growth history. This is the improved entropy familiar from renormalization-group treatments of black holes~\cite{Reuter1998,Percacci2017,Eichhorn2019}, where it is obtained by demanding consistency with the first law rather than postulated.

The identity is not vacuous, since the alternative prescription in which the coupling is simply evaluated at the horizon fails. For the Tsallis-Cirto entropy one finds $A/[4G_{\rm TC}(r_+)]=\delta\,S_{\rm TC}(r_+)$, and for the R\'enyi entropy $A/[4G_R(r_+)]=\pi r_+^2/(1+\pi\lambda r_+^2)$, neither of which reproduces the input entropy. The construction thus selects the first-law-consistent form \eqref{eq:improved_area_law} and rejects the inconsistent one.

The matter-sector reading can be excluded independently of any statement about Wald functionals. If the geometry is regarded as a solution of Einstein gravity sourced by a dressed distribution, the total horizon entropy is the sum of the gravitational and matter contributions,
\begin{equation}
 S_{\rm tot}(r_+)=\frac{A}{4G_N}+S_{\rm matter}(r_+) \ ,
 \qquad S_{\rm matter}\geq0 \ ,
 \label{eq:GSL_decomposition}
\end{equation}
so that consistency with the entropy input requires $S(r_+)\geq S_{\rm BH}(r_+)$ at every horizon radius. This inequality fails for two of the models considered here. For the R\'enyi entropy, concavity of the logarithm gives $\lambda^{-1}\ln(1+\lambda x)<x$ for all $x>0$ and all $\lambda>0$, with a deficit that grows without bound. For the Tsallis-Cirto entropy with $\delta<1$ one has $x^\delta<x$ for every $x>1$, that is for every macroscopic horizon. In both cases the deficit would have to be supplied by a negative matter entropy, which is a problem from a physical point of view. The remaining models are consistent with the inequality: the Kaniadakis and exponentially corrected entropies exceed the area law for all $x$, while the Barrow, logarithmically corrected and Tsallis-Cirto ($\delta>1$) entropies do so for $x>1$, falling below only for horizons smaller than the scale at which the entropy normalization is fixed, where the construction makes no claim in any case.

Three positions should therefore be distinguished. A local anisotropic fluid minimally coupled to Einstein gravity is excluded, both by the Wald argument and by Eq.~\eqref{eq:GSL_decomposition}. A covariantly nonlocal matter sector, in which the operator acting on the source involves $\Box$ and can generate curvature couplings through commutators of covariant derivatives, is not excluded by the Wald argument alone, since in that case the matter Lagrangian need not satisfy $\partial L_{\rm matter}/\partial R_{abcd}=0$; it remains, however, subject to Eq.~\eqref{eq:GSL_decomposition}. The modified gravitational sector, in which the form factor is carried by the gravitational action and the coupling runs, reproduces the input entropy exactly through Eq.~\eqref{eq:improved_area_law} for every model considered.

Two consequences follow. First, the entropy input breaks the degeneracy noted in Sec.~\ref{sec:nonlocal}: the generalized entropy is recovered in the reading in which the gravitational coupling runs, and not in the one in which the geometry is sourced by matter within unmodified Einstein gravity. Second, Eq.~\eqref{eq:improved_area_law} supplies the thermodynamic justification for the running coupling introduced in this section, which would otherwise rest on analogy alone. What is not established here is a Noether-charge computation in a fully covariant nonlocal theory, where Wald's construction is known to require care; Eq.~\eqref{eq:improved_area_law} should be read as strong evidence that such a computation would return the input entropy, not as a substitute for it. We also note that the thermodynamic statements above involve $G_S$ evaluated at the horizon only; the radial dependence of $G_S(r)$ and of $\beta_G^{(r)}(r)$ at $r\neq r_+$ is a geometric readout of the metric and carries no independent thermodynamic meaning.

The general relations derived above apply to any entropy functional that admits a differentiable cumulative profile. In the following sections we illustrate the formalism by considering several representative entropy models and explicitly constructing their associated effective matter distributions, nonlocal operators, momentum-space form factors, and running Newton couplings.
\section{Entropy models: operator, smeared mass, and geometry}\label{sec:examples}

Having established the general correspondence between entropy-generated geometries and effective nonlocal gravity, we now apply the formalism to several representative entropy models. For each case, the entropy function uniquely determines the associated nonlocal operator, which in turn generates the effective matter distribution, the cumulative mass profile, and the corresponding spacetime geometry. This organization makes the physical origin of each solution transparent and provides a unified description of geometries arising from different entropy corrections. The correspondence is summarized by the sequence
\begin{equation}
 S(r)
 \longrightarrow
 \mathcal A_S^{-2}
 \longrightarrow
 \rho_S(r)
 \longrightarrow
 m_S(r)
 \longrightarrow
 f_S(r) \ ,
 \label{eq:model_chain}
\end{equation}
where the entropy determines the effective nonlocal operator, the latter defines the dressed (or smeared) matter distribution, and the resulting cumulative mass generates the metric function through the Einstein equations.

Throughout this section, we employ the notation
\begin{equation}
 x\equiv S_{\rm BH}=\pi r^2, \qquad m_S(r)=M u_S(r) \ ,
 \label{eq:notation_examples}
\end{equation}
where $u_S(r)$ denotes the cumulative mass profile already defined in Eq.~\eqref{eq:u_def} associated with the entropy model.

Restoring Newton's constant, the metric function for the static, spherically symmetric line element reads
\begin{equation}
 f_S(r)= 1-\frac{2G_Nm_S(r)}{r} = 1-\frac{4\pi G_NM}{S'(r)} \ ,
 \label{eq:master_metric_reorganized}
\end{equation}
where the effective Newton coupling is
\begin{equation}
 G_S(r)=G_Nu_S(r) \ .
\end{equation}
The event horizon is determined by the condition $f_S(r_h)=0$, which is equivalently expressed as
\begin{equation}
 r_h = 2G_Nm_S(r_h) \qquad\Longleftrightarrow\qquad S'(r_h) = 4\pi G_NM \ .
 \label{eq:master_horizon_reorganized}
\end{equation}
The conditions established in Sec.~\ref{sec:entropy_geometry}, namely $S'>0$ and $S''>0$, are satisfied for all $r>0$ by the Bekenstein-Hawking, Barrow, Kaniadakis, exponentially corrected (with $0<\eta<1$) and LQG (with $q<1$) entropies, which are therefore unrestricted. For the R\'enyi, Tsallis-Cirto and logarithmically corrected entropies the convexity condition \eqref{eq:convexity_condition} holds only on part of the radial domain, and the corresponding restrictions are stated in the relevant subsections below.

Before discussing specific entropy models, it is useful to establish the general relation between the reconstructed nonlocal operator and the cumulative mass profile. We begin with a static point source whose energy density is dressed by the inverse nonlocal form factor,
\begin{equation}
 \rho_S(\mathbf r) = M\,\mathcal A_S^{-2}(-\nabla^2)\delta^{(3)}(\mathbf r) \ ,
 \label{eq:rho_operator_master}
\end{equation}
where $\mathcal A_S^{-2}$ denotes the effective nonlocal operator associated with the entropy model under consideration.

Employing the Fourier representation of the three-dimensional Dirac delta function,
\begin{equation}
 \delta^{(3)}(\mathbf r)
 =
 \int\frac{\dd^3k}{(2\pi)^3}\,
 e^{i\mathbf k\cdot\mathbf r},
 \label{eq:delta_fourier_master}
\end{equation}
the effective energy density can be expressed as
\begin{equation}
 \frac{\rho_S(r)}{M}
 =
 \int\frac{\dd^3k}{(2\pi)^3}\,
 \mathcal A_S^{-2}(k)\,
 e^{i\mathbf k\cdot\mathbf r}.
\end{equation}
For a rotationally invariant form factor, the angular integration may be performed explicitly, yielding
\begin{equation}
 \frac{\rho_S(r)}{M}
 =
 \frac{1}{2\pi^2r}
 \int_0^\infty
 \dd k\,k\,
 \mathcal A_S^{-2}(k)
 \sin(kr).
 \label{eq:rho_radial_fourier}
\end{equation}

The corresponding mass enclosed within a sphere of radius $r$ is obtained by integrating the effective density,
\begin{align}
 m_S(r)
 &=
 4\pi
 \int_0^r
 \dd x\,x^2
 \rho_S(x) = M\,u_S(r) \ ,
 \label{eq:mass_from_operator_master}
\end{align}
which defines the dimensionless cumulative profile $u_S(r)$. Substituting Eq.~\eqref{eq:rho_radial_fourier} into the above expression gives
\begin{equation}
 u_S(r)
 =
 \frac{2}{\pi}
 \int_0^\infty
 \dd k\,
 \mathcal A_S^{-2}(k)
 \frac{\sin(kr)-kr\cos(kr)}{k},
 \label{eq:u_from_operator_master}
\end{equation}
thereby establishing the direct correspondence between the nonlocal operator in momentum space and the cumulative mass profile.

An equivalent representation follows directly in coordinate space. If the kernel
$\mathcal A_S^{-2}(-\nabla^2)\delta^{(3)}(\mathbf r)$
is known explicitly, then
\begin{equation}
 u_S(r)
 =
 4\pi
 \int_0^r
 \dd x\,x^2\,
 \mathcal A_S^{-2}(-\nabla^2)
 \delta^{(3)}(\mathbf x),
 \label{eq:u_from_kernel_master}
\end{equation}
which may be regarded as the operator definition of the cumulative profile.

For kernels containing an unresolved point contribution, the distributional part must be retained explicitly. Writing
\begin{equation}
 \mathcal A_S^{-2}(-\nabla^2)\delta^{(3)}(\mathbf r)
 =
 u_S(0)\,
 \delta^{(3)}(\mathbf r)
 +
 \frac{u'_S(r)}{4\pi r^2},
 \label{eq:kernel_u_master}
\end{equation}
one immediately finds
\begin{eqnarray}
 u_S(r) &=& 4\pi \int_0^r \dd x\,x^2\, \mathcal A_S^{-2}(-\nabla^2) \delta^{(3)}(\mathbf x) \nonumber \\
 &=& u_S(0)+ \int_0^r \dd x\,u'_S(x)
 \label{eq:kernel_integrates_to_u}
\end{eqnarray}
confirming the consistency of the coordinate-space representation.

The construction, therefore, follows the sequence
\begin{equation}
 \mathcal A_S^{-2}\rightarrow \rho_S(r) \rightarrow u_S(r) \rightarrow m_S(r) \rightarrow f_S(r) \ ,
 \label{eq:operator_to_geometry_chain}
\end{equation}
where the nonlocal operator determines the effective matter distribution, which in turn fixes the cumulative mass profile and the resulting spacetime geometry. Finally, the entropy-geometry correspondence provides the identification through Eq.~\eqref{eq:u_def}. Consequently, each entropy model admits two complementary constructions. One may determine the effective nonlocal kernel starting from the entropy function $S(r)$, or conversely reconstruct the cumulative profile from the kernel through Eq.~\eqref{eq:u_from_kernel_master}. An agreement between the two procedures provides a nontrivial consistency check of the entropy-nonlocal gravity correspondence.

\subsection{Bekenstein-Hawking entropy}

We begin with the standard Bekenstein-Hawking entropy,
\begin{equation}
 S_{\rm BH}=x=\pi r^2 \ ,
\end{equation}
which corresponds to the classical area law of black-hole thermodynamics~\cite{Bekenstein1973,Hawking1975}. From Eq.~\eqref{eq:u_def}, the cumulative profile is simply
\begin{equation*}
 u_{\rm BH}(r)=1 \ .
\end{equation*}
Consequently, the effective Newton coupling remains constant,
\begin{equation*}
 G_{\rm BH}(r)=G_N \ ,
\end{equation*}
indicating the absence of any nonlocal modification.

The reconstructed operator as in Eq.~\eqref{eq:kernel_u_master} therefore reduces to the identity,
\begin{equation*}
 \mathcal A_{\rm BH}^{-2}(-\nabla^2)\delta^{(3)}(\mathbf r) = \delta^{(3)}(\mathbf r) \ ,
\end{equation*}
or equivalently,
\begin{equation*}
 \mathcal A_{\rm BH}^{-2}(k) = \mathcal A_{\rm BH}^{2}(k) = 1 \ .
\end{equation*}
Thus, the matter source remains a point particle, and no smearing of the energy density occurs.

Since the inverse operator is the identity, Eq.~\eqref{eq:kernel_integrates_to_u} immediately gives
\begin{equation*}
 u_{\rm BH}(r) = 4\pi \int_0^r \dd x\,x^2 \delta^{(3)}(\mathbf x) = 1 \ ,
\end{equation*}
for every sphere enclosing the origin. Accordingly, 
\begin{equation*}
 m_{\rm BH}(r)=M \ ,
\end{equation*}
showing that the enclosed mass is independent of the radial coordinate. Finally, substituting the above result into Eq.~\eqref{eq:master_metric_reorganized} yields the form of metric function as
\begin{equation}
 f_S^{\rm BH}(r)=1-\frac{2G_NM}{r} \ .
 \label{eq:metric_BH}
\end{equation}
This is precisely the Schwarzschild solution~\cite{Schwarzschild1916}, representing the local limit of the present construction. The spacetime is asymptotically flat and is sourced by a point mass located at the origin, leading to the familiar curvature singularity at $r=0$. The Bekenstein-Hawking entropy, therefore, corresponds to the trivial nonlocal operator, providing the reference configuration against which the entropy-modified geometries discussed below may be compared.


\subsection{R\'enyi entropy}

We first consider the R\'enyi entropy, which constitutes one of the simplest nonextensive generalizations of the Bekenstein-Hawking area law and has found several applications in black-hole thermodynamics~\cite{Renyi1961, Czinner2016, Biro2013}. It is given by
\begin{equation}
S_R(x)=\frac{1}{\lambda}\ln(1+\lambda x) \ ,
\label{eq:renyi_entropy}
\end{equation}
where $\lambda$ denotes the nonextensive deformation parameter. In the limit $\lambda\rightarrow0$, Eq.~\eqref{eq:renyi_entropy} reduces continuously to the Bekenstein-Hawking entropy. Differentiating Eq.~\eqref{eq:renyi_entropy} with respect to the radial coordinate gives
\begin{align}
S_R'(r)&=\frac{2\pi r}{1+\pi\lambda r^2} \ . \nonumber
\end{align}
Using the entropy-geometry correspondence established in Eq.~\eqref{eq:u_def}, the corresponding cumulative mass profile becomes
\begin{equation}
u_R(r) = 1+\pi\lambda r^2 \ .
\label{eq:renyi_u}
\end{equation}

The effective nonlocal kernel as in Eq.~\eqref{eq:kernel_u_master} is
\begin{align}
\mathcal A_R^{-2}(-\nabla^2)\delta^{(3)}(\mathbf r) &=\delta^{(3)}(\mathbf r)+\frac{\lambda}{2r} \ .
\label{eq:renyi_kernel_x}
\end{align}
Thus, in contrast to the Bekenstein-Hawking case, the point source acquires an additional long-range contribution generated by the nonlocal operator.

To determine the momentum-space form factor, we employ the standard Fourier-transform identities as
\begin{equation*}
\mathcal F\!\left[\delta^{(3)}(\mathbf r)\right]=1 \ ,
\qquad
\mathcal F\!\left[\frac{1}{r}\right]
=
\frac{4\pi}{k^2} \ ,
\end{equation*}
where the second identity follows from the Green function of the three-dimensional Laplacian~\cite{Jackson}. Taking the Fourier transform of Eq.~\eqref{eq:renyi_kernel_x} yields
\begin{align}
\mathcal A_R^{-2}(k)&=1+\frac{\lambda}{2}\frac{4\pi}{k^2}  = 1+\frac{2\pi\lambda}{k^2} \ . \nonumber 
\end{align}
The corresponding form factor is therefore
\begin{equation*}
\mathcal A_R^{2}(k)=\frac{1}{\mathcal A_R^{-2}(k)}=\frac{k^2}{k^2+2\pi\lambda} \ ,
\end{equation*}
or equivalently,
\begin{equation*}
\mathcal A_R^{-2}(-\nabla^2) = 1 +2\pi\lambda(-\nabla^2)^{-1} \ .
\end{equation*}
The nonlocal modification is thus governed by the inverse Laplacian, indicating that the R\'enyi entropy generates an infrared dressing of the matter source.

As a consistency check, the cumulative profile can be reconstructed directly from the operator. Substituting Eq.~\eqref{eq:renyi_kernel_x} into the operator definition in Eq.~\eqref{eq:kernel_integrates_to_u} gives
\begin{align}
u_R(r) &=4\pi \int_0^r \dd x\,x^2 \left[\delta^{(3)}(\mathbf x)+\frac{\lambda}{2x}\right] =1+\pi\lambda r^2 \ , \nonumber 
\end{align}
which exactly reproduces Eq.~\eqref{eq:renyi_u}. Accordingly, the corresponding cumulative mass profile is
\begin{equation}
m_R(r) = M\,u_R(r) = M\left(1+\pi\lambda r^2\right) \ .
\label{eq:renyi_mass}
\end{equation}

The associated spacetime geometry follows from the master relation as in Eq.~\eqref{eq:master_metric_reorganized} and substituting Eq.~\eqref{eq:renyi_mass} yields
\begin{align}
f_R(r) &= 1-\frac{2G_NM}{r}\left(1+\pi\lambda r^2\right) \ .
\label{eq:f_Renyi_geometry}
\end{align}
This exactly matches the metric function derived in~\cite{Anand2025} and reduces to the Schwarzschild solution in the limit $\lambda \rightarrow 0$. The corresponding running Newton coupling is
\begin{equation*}
G_R(r) = G_Nu_R(r) = G_N\left(1+\pi\lambda r^2\right) \ .
\end{equation*}

Finally, we note that the cumulative profile grows quadratically, $u_R(r)\sim r^2$, implying that the enclosed mass increases without bound, $m_R(r)\sim r^2$. Consequently, the metric function contains a term linear in $r$, and the resulting spacetime is neither asymptotically Schwarzschild nor asymptotically flat. The R\'enyi solution should therefore be regarded as an effective infrared geometry, valid within a finite radial domain or supplemented by an appropriate large-distance completion.

This is made precise by the criterion \eqref{eq:convexity_condition}. One has $S_R''(r)>0$ only for $r<1/\sqrt{\pi\lambda}$, the marginal radius being an extremum of $S'_R$ and hence, by Eq.~\eqref{eq:T_general}, a degenerate horizon. The corresponding extremal mass
\begin{equation*}
 M_{\rm ext}=\frac{1}{4G_N\sqrt{\pi\lambda}}
\end{equation*}
is an \emph{upper} bound: for $M>M_{\rm ext}$ the metric function has no zero, while for $M<M_{\rm ext}$ it has two, the inner one being the black-hole horizon with $T_H>0$ and the outer one a cosmological horizon, as expected from the term linear in $r$ in Eq.~\eqref{eq:f_Renyi_geometry}. The thermodynamic reconstruction is thus underwritten only on $r<1/\sqrt{\pi\lambda}$; beyond that radius the geometry remains well defined but is obtained by analytic continuation.

\subsection{Tsallis-Cirto entropy}

We next consider the Tsallis-Cirto entropy, which provides a nonadditive generalization of the Bekenstein-Hawking area law and has been widely employed in generalized black-hole thermodynamics and cosmology~\cite{Tsallis1988, Cirto2013, Tsallis2013} and have the form
\begin{equation}
S_{\rm TC}(x)=x^\delta \ ,
\label{eq:tsallis_entropy}
\end{equation}
where $\delta$ is the nonextensive deformation parameter. The Bekenstein-Hawking entropy is recovered continuously in the limit $\delta\rightarrow1$. Differentiating Eq.~\eqref{eq:tsallis_entropy} with respect to the radial coordinate and employing the entropy-geometry correspondence established in Eq.~\eqref{eq:u_def}, the corresponding cumulative profile becomes
\begin{equation}
u_{\rm TC}(r) = \frac{1}{\delta}x^{1-\delta} = \frac{\pi^{1-\delta}}{\delta} r^{2-2\delta} \ .
\label{eq:tsallis_u}
\end{equation}
For $\delta=1$, one immediately recovers the Schwarzschild result, $u_{\rm TC}(r)=1$.

The corresponding effective energy density follows from the general relation
\begin{equation}
\rho(r)= \frac{M}{4\pi r^2} \frac{du(r)}{dr}= M\frac{1-\delta}{2\delta\pi^\delta} r^{-2\delta-1} \ .
\end{equation}

To determine the corresponding momentum-space form factor, we employ the Fourier transform of the Riesz kernel as
\begin{equation*}
\mathcal F \!\left[ r^{-\alpha}\right] = 2^{3-\alpha}\pi^{3/2}\frac{\Gamma\!\left(\frac{3-\alpha}{2}\right)}{\Gamma\!\left(\frac{\alpha}{2}\right)}k^{\alpha-3} \ ,
\end{equation*}
understood through analytic continuation whenever necessary. Substituting $\alpha=2\delta+1$ gives
\begin{equation}\label{eq:Cdelta}
\mathcal A_{\rm TC}^{-2}(k) = C_\delta\,k^{2\delta-2} \quad \text{where}\; C_\delta = \frac{2^{1-2\delta} \pi^{3/2-\delta}\Gamma(2-\delta)}{\delta\Gamma(\delta+\tfrac12)}\ .
\end{equation}
The corresponding coordinate-space operator is therefore
\begin{equation*}
\mathcal A_{\rm TC}^{-2}(-\nabla^2) = C_\delta (-\nabla^2)^{\delta-1} \ ,
\end{equation*}
or equivalently,
\begin{equation}
\mathcal A_{\rm TC}^{2}(-\nabla^2) = C_\delta^{-1} (-\nabla^2)^{1-\delta} \ .
\end{equation}
Thus, the Tsallis-Cirto deformation naturally gives rise to a fractional Laplace operator whose order is determined by the nonextensive parameter $\delta$. At $\delta=1$, one has $C_1=1$, recovering the local Einstein theory.

As a consistency check, the cumulative profile can be reconstructed directly from the operator. The Riesz kernel gives
\begin{equation*}
\mathcal A_{\rm TC}^{-2}(-\nabla^2) \delta^{(3)}(\mathbf r)  = \frac{1-\delta}{2\delta\pi^\delta}r^{-2\delta-1},
\qquad (r>0) \ ,
\end{equation*}
which coincides with the effective density kernel obtained above. Substituting this expression into Eq.~\eqref{eq:kernel_integrates_to_u} gives
\begin{align}
u_{\rm TC}(r) &= 4\pi \int_0^r dx\,x^2 \left[\frac{1-\delta}{2\delta\pi^\delta}x^{-2\delta-1}\right] = \frac{\pi^{1-\delta}}{\delta}r^{2-2\delta} \ , \nonumber 
\end{align}
where the integral converges at the origin for $\delta<1$ and is otherwise defined by analytic continuation in $\delta$, the singular endpoint contribution at $r=0$ being discarded. This exactly reproduces Eq.~\eqref{eq:tsallis_u}. Accordingly, the cumulative mass profile becomes
\begin{equation}
m_{\rm TC}(r)= M\,u_{\rm TC}(r) = \frac{M\pi^{1-\delta}} {\delta} r^{2-2\delta} \ .
\label{eq:tsallis_mass}
\end{equation}
The associated spacetime geometry follows from the relation in Eq.~\eqref{eq:master_metric_reorganized} by substituting Eq.~\eqref{eq:tsallis_mass} yields
\begin{align}
f_{\rm TC}(r) &= 1- \frac{2G_NM\pi^{1-\delta}}{\delta} r^{1-2\delta} \ . \nonumber
\end{align}
This expression exactly reproduces the metric function obtained in Ref.~\cite{Anand2025} and approaches the Schwarzschild solution in the limit $\delta\rightarrow 1$. The corresponding running Newton coupling is
\begin{equation*}
G_{\rm TC}(r) = G_Nu_{\rm TC}(r) = \frac{G_N\pi^{1-\delta}} {\delta} r^{2-2\delta} \ .
\end{equation*}

Finally, we note that the asymptotic behaviour depends sensitively on the value of the nonextensive parameter. For $\delta>1/2$, the gravitational correction decreases with radial distance, although it reproduces the Schwarzschild $1/r$ falloff only at $\delta=1$. In contrast, for $\delta<1/2$, the correction grows at large distances, indicating an intrinsically infrared-modified geometry. These features are governed by a single criterion: since $S''_{\rm TC}$ has the sign of $2\delta-1$, the convexity condition \eqref{eq:convexity_condition} holds for $\delta>1/2$ and fails for $\delta<1/2$ at every radius. In the latter case $f_{\rm TC}$ decreases monotonically from unity and its single zero is a cosmological-type horizon with $T_H<0$, so that the geometry describes no black hole at all. The special value $\delta=1/2$ corresponds to $S''\equiv0$, i.e. $T_H\equiv0$: the metric function is constant and the horizon, when it exists, is degenerate. Consequently, except for the Einstein limit $\delta=1$, the Tsallis-Cirto entropy describes a scale-free fractional modification of gravity rather than a conventional finite-mass Schwarzschild geometry.


\subsection{Barrow entropy}

We next consider the Barrow entropy, proposed to incorporate possible quantum-gravitational deformations of the black-hole horizon arising from a fractalization of the horizon geometry~\cite{Barrow2020, Saridakis2020}. It is given by
\begin{equation}
S_B(x)=x^{1+\Delta/2} , \qquad 0\le\Delta\le1 \ ,
\label{eq:barrow_entropy}
\end{equation}
where $\Delta$ denotes the Barrow deformation parameter. The standard Bekenstein-Hawking entropy is recovered continuously in the limit $\Delta\rightarrow0$.

Differentiating Eq.~\eqref{eq:barrow_entropy} with respect to the radial coordinate and using the entropy-geometry correspondence as in Eq.~\eqref{eq:u_def}, the corresponding cumulative profile becomes
\begin{equation}
u_B(r) = \frac{2}{2+\Delta} \pi^{-\Delta/2} r^{-\Delta} \ .
\label{eq:barrow_u}
\end{equation}
For $\Delta=0$, one immediately recovers the Schwarzschild result, $u_B(r)=1$. Using an idea similar to the Tsallis-Cirto case, the momentum-space form factor, we use the Fourier-transform identity
\begin{equation*}
\mathcal F \!\left[r^{-3-\Delta}\right]\propto k^\Delta \ ,
\end{equation*}
valid in the distributional sense for fractional powers. Consequently,
\begin{equation*}
\mathcal A_B^{-2}(k) = C_{1+\Delta/2} k^\Delta \ ,
\end{equation*}
where $C_{1+\Delta/2}$ can be computed using Eq.~\eqref{eq:Cdelta}. Passing to coordinate space gives
\begin{equation*}
\mathcal A_B^{-2}(-\nabla^2) = C_{1+\Delta/2}(-\nabla^2)^{\Delta/2} \ ,
\end{equation*}
showing that the Barrow deformation is equivalent to replacing the local Laplacian by a fractional Laplace operator of order $\Delta$.

Again, as a consistency check, we will reconstruct the cumulative profile directly from the operator using Eq.~\eqref{eq:kernel_integrates_to_u} as 
\begin{align}
&u_B(r)= 4\pi \int_0^r dx\,x^2 \left[ -\frac{\Delta}{2\pi(2+\Delta)}\pi^{-\Delta/2} x^{-3-\Delta}\right]
\nonumber\\
&=-\frac{2\Delta}{2+\Delta}\pi^{-\Delta/2}\int_0^r dx\,x^{-1-\Delta}=\frac{2}{2+\Delta} \frac{\pi^{-\Delta/2}}{ r^{\Delta}} \ ,
\end{align}
which exactly reproduces Eq.~\eqref{eq:barrow_u}. Furthermore, the corresponding cumulative mass profile becomes
\begin{equation}
m_B(r) = M\,u_B(r) = \frac{2M}{2+\Delta}\pi^{-\Delta/2}r^{-\Delta} \ .
\label{eq:barrow_mass}
\end{equation}
The associated spacetime geometry follows from Eq.~\eqref{eq:master_metric_reorganized} upon substituting Eq.~\eqref{eq:barrow_mass},
\begin{align}
f_B(r) &= 1- \frac{4G_NM}{2+\Delta}\pi^{-\Delta/2} r^{-1-\Delta} \ . \nonumber 
\end{align}
The resulting metric function is identical to that reported in~\cite{Anand2025}, while the Schwarzschild geometry is recovered as $\Delta \rightarrow 0$ and the corresponding running Newton coupling takes the form 
\begin{equation*}
G_B(r)= \frac{2G_N} {2+\Delta}\pi^{-\Delta/2}r^{-\Delta} \ .
\end{equation*}

Finally, we note that the cumulative profile decreases as
$u_B(r)\sim r^{-\Delta}$ for $\Delta>0$, implying that the effective enclosed mass also decreases with radial distance. Consequently, the gravitational potential falls off as $r^{-1-\Delta}$, faster than the Schwarzschild $1/r$ behavior. Although the metric approaches Minkowski spacetime asymptotically, the coefficient of the standard $1/r$ term vanishes, indicating that the geometry does not describe the exterior field of a finite ADM mass. Moreover, the power-law kernel remains singular at the origin, reflecting the intrinsically fractional and scale-dependent gravitational response generated by Barrow entropy.

\subsection{Kaniadakis entropy}

We next consider the Kaniadakis entropy, which constitutes another one-parameter deformation of the Bekenstein-Hawking area law motivated by the $\kappa$-deformed statistical framework introduced by Kaniadakis~\cite{Kaniadakis2001, Kaniadakis2002, Abreu2016}. It is given by
\begin{equation}
S_K(x) = \frac{\sinh(\kappa x)}{\kappa} \ ,
\label{eq:kani_entropy}
\end{equation}
where $\kappa$ denotes the deformation parameter. In the limit $\kappa\rightarrow0$, one recovers the standard Bekenstein-Hawking entropy.

Now, using the entropy-geometry correspondence established in Eq.~\eqref{eq:u_def} and differentiating Eq.~\eqref{eq:kani_entropy} with respect to the radial coordinate, the corresponding cumulative profile reads
\begin{equation}
u_K(r) = \sech(\kappa\pi r^2) \ .
\label{eq:kani_u}
\end{equation}
For $\kappa=0$, Eq.~\eqref{eq:kani_u} reduces to unity, i.e., the Schwarzschild result. The corresponding effective nonlocal kernel follows from Eq.~\eqref{eq:kernel_u_master} as
\begin{align}
\mathcal A_K^{-2}(-\nabla^2)\delta^{(3)}(\mathbf r) &= \delta^{(3)}(\mathbf r) \nonumber\\
&\quad -\frac{\kappa}{2r} \tanh(\kappa\pi r^2) \sech(\kappa\pi r^2) \ .
\label{eq:kani_kernel}
\end{align}
The Kaniadakis deformation generates a nonlinear, exponentially decaying correction to the point source, leading to a screened effective matter distribution.

The corresponding momentum-space form factor is obtained by taking the Fourier transform of Eq.~\eqref{eq:kani_kernel}. Using spherical symmetry, one finds
\begin{align}
\mathcal A_K^{-2}(k) &= 1 - 2\pi\kappa \int_0^\infty dr\,r\, \tanh(\kappa\pi r^2) \sech(\kappa\pi r^2) \frac{\sin kr}{kr} \ . \nonumber
\end{align}
In contrast to the previous examples, the resulting form factor is a non-polynomial function of the momentum and therefore does not admit a simple fractional-power representation.

As a consistency check, the cumulative profile may be reconstructed directly from the operator. Substituting Eq.~\eqref{eq:kani_kernel} into the general definition Eq.~\eqref{eq:kernel_integrates_to_u} gives
\begin{align}
&u_K(r) =4\pi \int_0^r dx\,x^2 \left[ \delta^{(3)}(\mathbf x)- \frac{\kappa}{2x}\tanh(\kappa\pi x^2)\sech(\kappa\pi x^2)\right] \nonumber\\
&= 1- 2\pi\kappa \int_0^r dx\,x \tanh(\kappa\pi x^2) \sech(\kappa\pi x^2) = \sech(\kappa\pi r^2) \ , \nonumber 
\end{align}
thereby reproducing Eq.~\eqref{eq:kani_u}. With the help of this, the cumulative mass profile becomes
\begin{equation}
m_K(r) = M\,u_K(r) = M\sech(\kappa\pi r^2) \ .
\label{eq:kani_mass}
\end{equation}
Finally, the associated spacetime geometry can be constructed using Eq.~\eqref{eq:master_metric_reorganized} by substituting Eq.~\eqref{eq:kani_mass} results in
\begin{align}
f_K(r)&= 1- \frac{2G_NM}{r} \sech(\kappa\pi r^2) \ . \nonumber
\end{align}
This result is fully consistent with the metric function derived in~\cite{Anand2025} and smoothly reduces to the Schwarzschild solution in the limit $\kappa \rightarrow 0$.
The corresponding running Newton coupling is $G_N\sech(\kappa\pi r^2)$.

Near the origin, the metric function admits the expansion
\begin{equation*}
f_K(r) = 1 - \frac{2G_NM}{r} + G_NM\kappa^2\pi^2r^3 +\mathcal O(r^7) \ ,
\end{equation*}
showing that the Schwarzschild curvature singularity persists despite the entropy deformation. On the other hand, the cumulative profile satisfies
\begin{equation*}
u_K(r)\sim 2e^{-\kappa\pi r^2},
\qquad (r\rightarrow\infty) \ ,
\end{equation*}
so that the effective mass is exponentially screened at large distances. Consequently, the metric approaches Minkowski spacetime faster than any inverse power of the radial coordinate, and the coefficient of the conventional Schwarzschild $1/r$ term vanishes asymptotically. The resulting geometry should therefore be interpreted as describing a compensated nonlocal source rather than the exterior field of a finite ADM mass.

\subsection{Logarithmically corrected entropy}

Now, we consider the logarithmically corrected entropy, which naturally arises from quantum corrections to the Bekenstein-Hawking area law in several approaches to quantum gravity, including loop quantum gravity and quantum geometry~\cite{KaulMajumdar2000, Carlip2000, Sen2013}. The form of the logarithmically corrected entropy is
\begin{equation}
S_{\log}(r) = x + \alpha \ln{x} \ ,
\label{eq:log_entropy_final}
\end{equation}
where $\alpha$ parametrizes the strength of the logarithmic correction. In the limit $\alpha\rightarrow0$, one recovers the standard Bekenstein-Hawking entropy.

Differentiation of the logarithmically corrected entropy, Eq.~\eqref{eq:log_entropy_final}, with respect to \(r\) gives
\begin{equation}
S'_{\log}(r) =2\pi r + \frac{2\alpha}{r} = \frac{2(\pi r^2+\alpha)}{r} \ .
\end{equation}
Using Eq.~\eqref{eq:u_def}, the corresponding cumulative distribution function is then obtained as
\begin{equation}
u_{\log}(r) = \frac{\pi r^2}{\pi r^2+\alpha},
\label{eq:log_u_final}
\end{equation}
which continuously approaches the Schwarzschild value in the limit $\alpha\rightarrow0$. Accordingly, the cumulative mass profile is
\begin{equation}
m_{\log}(r) = M \frac{\pi r^2}{\pi r^2+\alpha} \ .
\label{eq:log_mass_final}
\end{equation}

The corresponding effective nonlocal kernel follows from the general relation in Eq.~\eqref{eq:kernel_u_master},
\begin{equation*}
\mathcal A_{\log}^{-2}(-\nabla^2)\delta^{(3)}(\mathbf r)=u_{\log}(0)\,\delta^{(3)}(\mathbf r)+\frac{u'_{\log}(r)}{4\pi r^2} \ .
\end{equation*}
For positive $\alpha$, one has
\begin{equation}
u_{\log}(0)=0 \ ,
\end{equation}
so that the Dirac delta contribution disappears completely. Differentiating Eq.~\eqref{eq:log_u_final} gives
\begin{equation*}
u'_{\log}(r) = \frac{2\pi\alpha r}{(\pi r^2+\alpha)^2} \ ,
\end{equation*}
and therefore
\begin{equation*}
\frac{\rho_{\log}(r)}{M} =\mathcal A_{\log}^{-2}(-\nabla^2)\delta^{(3)}(\mathbf r) = \frac{\alpha}{2r(\pi r^2+\alpha)^2} \ .
\end{equation*}
Unlike the previous examples, the logarithmic correction completely replaces the point source with a smooth extended distribution. As a consistency check, the cumulative profile may be reconstructed directly from the operator using Eq.~\eqref{eq:kernel_integrates_to_u},

\begin{align}
u_{\log}(r) &= 4\pi \int_0^r dx\,x^2 \frac{\alpha} {2x(\pi x^2+\alpha)^2}= \frac{\pi r^2}{\pi r^2+\alpha} \ , \nonumber
\end{align}
thereby reproducing Eq.~\eqref{eq:log_u_final} exactly. The corresponding momentum-space form factor is defined through the spherical Fourier transform,
\begin{eqnarray}
\mathcal A_{\log}^{-2}(k) &=& 4\pi \int_0^\infty dr\,r^2 \frac{\alpha}{2r(\pi r^2+\alpha)^2}\frac{\sin kr}{kr} \ , \nonumber
\end{eqnarray}
which provides the nonlocal operator associated with the logarithmic entropy. The Fourier transform integral does not admit a simple closed-form expression in terms of elementary functions. Symbolic evaluation in \textit{Mathematica} expresses the result in terms of Meijer G-functions\footnote{$\mathcal A_{\log}^{-2}(k)=\frac{\pi  G_{1,3}^{2,1}\left(\frac{k^2 \alpha }{4 \pi }|
\begin{array}{c}
 \frac{1}{2} \\
 \frac{1}{2},\frac{3}{2},0 \\
\end{array}
\right)}{\sqrt{\alpha } k}$}, whose explicit form is not particularly illuminating. Consequently, we do not reproduce the lengthy analytical expression here, and instead, one can analyze the integral numerically whenever required.

Finally, the metric function associated with the logarithmically corrected entropy follows from Eq.~\eqref{eq:master_metric_reorganized} upon substituting Eq.~\eqref{eq:log_mass_final},

\begin{equation}
f_{\log}(r) = 1- \frac{2G_NM}{r}\frac{\pi r^2}{\pi r^2+\alpha} \ .
\end{equation}
The obtained metric function coincides with the expression presented in~\cite{Anand2025} and recovers the Schwarzschild spacetime in the limit $\alpha \rightarrow 0$, and the corresponding running Newton coupling is
\begin{equation*}
G_{\log}(r) = G_N \frac{\pi r^2}{\pi r^2+\alpha} \ .
\end{equation*}

At large radial distances,
\begin{equation*}
u_{\log}(r) = 1 -\frac{\alpha}{\pi r^2}+\mathcal O(r^{-4}) \ ,
\end{equation*}
and consequently the Schwarzschild geometry is recovered asymptotically. Near the origin,
\begin{equation*}
u_{\log}(r) = \frac{\pi}{\alpha}r^2 + \mathcal O(r^4) \ ,
\end{equation*}
which gives
\begin{equation*}
f_{\log}(r) = 1 - \frac{2\pi G_NM}{\alpha}r +\mathcal O(r^3) \ .
\end{equation*}
Although the logarithmic correction removes the Dirac delta contribution for $\alpha>0$, the effective density still behaves as $\rho_{\log}\propto1/r$ near the origin. Consequently, the solution should be interpreted as an entropy-generated extended source rather than a completely regular black-hole core.

The convexity condition \eqref{eq:convexity_condition} is satisfied here for $r>\sqrt{\alpha/\pi}$, the marginal radius being the minimum of $S'_{\log}$. The horizon structure is accordingly of Reissner-Nordstr\"om type,
\begin{equation*}
 r_\pm=G_NM\pm\sqrt{G_N^2M^2-\frac{\alpha}{\pi}} \ ,
\end{equation*}
with the event horizon $r_+>\sqrt{\alpha/\pi}$ always lying in the region where $T_H>0$, and the inner root a Cauchy horizon. In contrast with the R\'enyi case, the extremal mass
\begin{equation*}
 M_{\rm ext}=\frac{1}{G_N}\sqrt{\frac{\alpha}{\pi}}
\end{equation*}
is a \emph{lower} bound, below which the logarithmic correction removes the horizon altogether. 
\subsection{Exponentially corrected entropy}

As a further example, we consider an exponentially corrected entropy~\cite{ Medved:2004mh, Nojiri:2006be, Chatterjee:2020iuf}, which has been proposed as a phenomenological quantum correction to the Bekenstein–Hawking area law and introduces short-distance modifications while preserving the classical entropy at large horizon area and is given by
\begin{equation}
S_{\rm exp}(x) = x+\eta e^{-x} \ ,
\label{eq:exp_entropy}
\end{equation}
where $\eta$ is the exponential deformation parameter and in the limit $\eta\rightarrow0$, Eq.~\eqref{eq:exp_entropy} reduces continuously to the Bekenstein–Hawking entropy. Throughout we restrict to $0<\eta<1$, for which $S'_{\rm exp}(r)>0$ for all $r>0$; for $\eta\ge1$ the entropy derivative vanishes at $r_0=\sqrt{\ln\eta/\pi}$ and the reconstruction breaks down there. The radial derivative of the entropy expression in Eq.~\eqref{eq:exp_entropy} is readily evaluated as
\begin{equation*}
S'_{\rm exp}(r)=2\pi r \left(1-\eta e^{-\pi r^2}\right) \ .
\end{equation*}
Inserting this expression into the entropy-geometry mapping defined by Eq.~\eqref{eq:u_def} leads directly to the cumulative profile
\begin{equation}
u_{\rm exp}(r)= \frac{1}{1-\eta e^{-\pi r^2}} \ .
\label{eq:exp_u}
\end{equation}
For the limiting case, one immediately recovers the Schwarzschild result. Now, the cumulative mass profile is
\begin{equation}
m_{\rm exp}(r) = \frac{M}{1-\eta e^{-\pi r^2}} \ .
\label{eq:exp_mass}
\end{equation}

The corresponding effective nonlocal kernel follows from the general relation Eq.~\eqref{eq:kernel_u_master} as
\begin{equation*}
\mathcal A_{\rm exp}^{-2}(-\nabla^2) \delta^{(3)}(\mathbf r)=u_{\rm exp}(0)\,\delta^{(3)}(\mathbf r)+\frac{u'_{\rm exp}(r)}{4\pi r^2} \ .
\end{equation*}
Since $u_{\rm exp}(0)=(1-\eta)^{-1}$, the point source survives with a renormalized strength. Differentiating
Eq.~\eqref{eq:exp_u}, one obtains
\begin{equation*}
\frac{\rho_{\rm exp}(r)}{M}=\frac{\delta^{(3)}(\mathbf r)}{1-\eta}-\frac{\eta e^{-\pi r^2}}{2r\left(1-\eta e^{-\pi r^2}\right)^2} \ .
\end{equation*}
Unlike the logarithmically corrected entropy, the Dirac delta contribution is not removed but is instead rescaled by the factor $(1-\eta)^{-1}$, accompanied by an exponentially localized dressing cloud. For completeness, the cumulative profile can also be verified directly within the operator formalism. Using Eq.~\eqref{eq:kernel_integrates_to_u}, 
\begin{align}
u_{\rm exp}(r)&=4\pi\int_0^r dx\,x^2 \left[\frac{1}{1-\eta}\delta^{(3)}(\mathbf x)-\frac{\eta e^{-\pi x^2}}{2x\left(1-\eta e^{-\pi x^2}\right)^2} \right] \nonumber\\
&=
\frac{1}{1-\eta}-2\pi\eta \int_0^r dx\, \frac{x e^{-\pi x^2}}{\left(1-\eta e^{-\pi x^2}\right)^2} \ .
\end{align}
Using the identity
\begin{equation}
\frac{d}{dr}\left[\frac{1}{1-\eta e^{-\pi r^2}}\right] = -\frac{2\pi\eta r e^{-\pi r^2}}{\left(1-\eta e^{-\pi r^2}\right)^2} \ ,
\end{equation}
the result exactly reproduces Eq.~\eqref{eq:exp_u}.

The corresponding momentum-space form factor is defined through the spherical Fourier transform
\begin{equation*}
\mathcal A_{\rm exp}^{-2}(k) = \frac{1}{1-\eta}- 2\pi\eta \int_0^\infty dr\,r \frac{e^{-\pi r^2}}{\left(1-\eta e^{-\pi r^2}\right)^2}\frac{\sin kr}{kr} \ ,
\label{eq:exp_operator}
\end{equation*}
which uniquely specifies the associated nonlocal operator. Owing to the exponential profile, the transform does not admit a simple elementary closed form. Finally, we compute the form of associated spacetime geometry metric function using Eq.~\eqref{eq:master_metric_reorganized} by substituting Eq.~\eqref{eq:exp_mass} results in
\begin{equation*}
f_{\rm exp}(r) = 1- \frac{2G_NM}{r\left(1-\eta e^{-\pi r^2}\right)} \ .
\end{equation*}
The derived metric function is in exact agreement with~\cite{Anand2025}. Furthermore, the Schwarzschild geometry is recovered in the limiting case $\eta \rightarrow 0$. The running Newton coupling takes the form as
\begin{equation*}
G_{\rm exp}(r)= \frac{G_N}{1-\eta e^{-\pi r^2}} \ .
\end{equation*}

Near the origin,
\begin{equation*}
u_{\rm exp}(r) = \frac{1}{1-\eta}-\frac{\eta\pi}{(1-\eta)^2} r^2+ \mathcal O(r^4) \ ,
\end{equation*}
so that
\begin{equation*}
f_{\rm exp}(r)=1-\frac{2G_NM}{(1-\eta)r}+\mathcal O(r) \ .
\end{equation*}
Thus, the Schwarzschild curvature singularity is preserved, although its strength is rescaled by the deformation parameter. At large radial distances,
\begin{equation*}
u_{\rm exp}(r) = 1 + \eta e^{-\pi r^2} + \mathcal O(e^{-2\pi r^2}) \ ,
\end{equation*}
and consequently
\begin{equation*}
f_{\rm exp}(r) = 1- \frac{2G_NM}{r} +\mathcal O\!\left(\frac{e^{-\pi r^2}}{r} \right) \ .
\end{equation*}
The exponential correction therefore produces a localized short-distance modification of the Schwarzschild geometry while preserving the standard asymptotic mass. In contrast to the logarithmically corrected entropy, the point source remains present with a renormalized strength, whereas the accompanying exponential dressing rapidly vanishes at large distances.


\subsection{Loop Quantum Gravity (LQG) Entropy}

As a final example we consider an important class of entropy corrections that emerges in Loop Quantum Gravity (LQG), where the microscopic counting of horizon states can be combined with nonextensive statistical mechanics to yield a generalized black-hole entropy~\cite{Nojiri:2022ljp, Czinner:2015eyk}. In the present framework, this entropy provides another illustrative example of how generalized horizon thermodynamics can be translated into an effective nonlocal gravitational description. Unlike the algebraic modifications generated by the Tsallis--Cirto or Barrow entropies, the LQG entropy considered here gives rise to an exponential profile, leading to a Gaussian-type dressing of the gravitational source. The entropy is given by
\begin{equation}
S_{\rm LQG}(A) = \frac{1}{1-q}\exp\!\left[(1-q)\Lambda(\gamma_0)x-1\right],
\label{eq:LQG_entropy}
\end{equation}
where $q$ denotes the nonextensive parameter, and $\Lambda(\gamma_0)=\tfrac{\ln2}{\sqrt3\,\pi\gamma_0}$,
with $\gamma_0$ representing the Barbero-Immirzi parameter. For convenience, Eq.~\eqref{eq:LQG_entropy} may be written directly as a function of the radial coordinate,
\begin{equation}
S_{\rm LQG}(r)=\frac{e^{-1}}{1-q}\exp\!\left[(1-q)\Lambda\pi r^2\right] \ .
\label{eq:LQG_entropy_r}
\end{equation}

Using the entropy–geometry correspondence introduced in Sec.~\ref{sec:entropy_geometry}, the cumulative profile requires the derivative of the entropy with respect to $r$, which immediately yields
\begin{align}
u_{\rm LQG}(r) = \frac{e}{\Lambda}\exp\!\left[-(1-q)\Lambda\pi r^2\right] \ .
\label{eq:LQG_u}
\end{align}
The cumulative profile therefore decreases exponentially with the radial coordinate, indicating that the entropy deformation produces a localized gravitational dressing rather than the power-law behaviour encountered for several of the previous entropy models. Since the enclosed mass is related to the cumulative profile through
\begin{equation}
m(r)=Mu(r),
\end{equation}
the corresponding cumulative mass function becomes
\begin{equation}
m_{\rm LQG}(r)=\frac{Me}{\Lambda}\exp\!\left[-(1-q)\Lambda\pi r^2 \right] \ .
\label{eq:LQG_mass}
\end{equation}

The effective source generated by the entropy deformation follows directly from the kernel relation Eq.~\eqref{eq:kernel_u_master},
\begin{align}
\frac{\rho_{\rm LQG}(r)}{M}&=\frac{e}{\Lambda}\delta^{(3)}(\mathbf r)-\frac{e(1-q)\exp\!\left[(q-1)\Lambda\pi r^2\right]}{2r} \ .
\label{eq:LQG_density}
\end{align}

Equation~\eqref{eq:LQG_density} clearly exhibits the two distinct components of the effective source. The first is a renormalized point contribution located at the origin, while the second is an exponentially localized cloud that dresses the point source. As a consistency check, the cumulative profile can be reconstructed directly from the effective kernel using the inverse relation in Eq.~\eqref{eq:u_from_kernel_master}, one finds
{\small
\begin{align}
\frac{u_{\rm LQG}(r)}{4\pi}=\int_0^r dx\,\left[\frac{e\,x^2\,\delta^{(3)}(\mathbf{x})}{\Lambda}-\frac{x^2e(1-q)\exp\!\left((q-1)\Lambda\pi x^2\right)}{2x}\right] \ , \nonumber
\end{align}}
by doing this integration one exactly reproduces Eq.~\eqref{eq:LQG_u}. 

Now, the corresponding momentum-space form factor is obtained by taking the spherical Fourier transform of the kernel and introducing $a=(1-q)\Lambda\pi$ the required integral becomes
\begin{equation}
\int_0^\infty dr\, \exp(-ar^2) \frac{\sin kr}{k}=\frac{1}{k\sqrt{a}}\,\mathrm{DawsonF}\!\left(\frac{k}{2\sqrt{a}}\right) \ ,
\end{equation}
where $\mathrm{DawsonF}(x)$ denotes Dawson's integral and the condition $\Re(a)>0$ must be satisfied. Consequently, the momentum-space operator assumes the exact closed form as
\begin{equation}
\mathcal A_{\rm LQG}^{-2}(k)=\frac{e}{\Lambda}-\frac{2\pi e(1-q)}{k\sqrt{\pi(1-q)\Lambda}}\,\mathrm{DawsonF}\!\left(\frac{k}{2\sqrt{\pi(1-q)\Lambda}}\right) \ .
\label{eq:LQG_formfactor_closed}
\end{equation}
This expression completely characterizes the effective nonlocal operator associated with the LQG entropy deformation. 

Having reconstructed the nonlocal operator, the corresponding spacetime geometry follows immediately from the relation as in Eq.~\eqref{eq:master_metric_reorganized}
\begin{equation}
f_{\rm LQG}(r)=1-\frac{2G_NMe}{\Lambda r}\exp\!\left[-(1-q)\Lambda\pi r^2 \right] \ ,
\label{eq:LQG_metric}
\end{equation}
which coincides with the entropy-generated geometry obtained directly from the entropy-geometry correspondence in Ref.~\cite{Anand2025}. Within the operator interpretation developed here, the cumulative profile simultaneously determines the scale dependence of the gravitational coupling. Using this one finds
\begin{equation}
G_{\rm LQG}(r) = \frac{G_Ne}{\Lambda} \exp\!\left[-(1-q)\Lambda\pi r^2 \right] \ .
\end{equation}
The gravitational coupling therefore decreases exponentially with the radial coordinate whenever $(1-q)\Lambda>0$, illustrating that the entropy deformation induces a scale-dependent screening of the gravitational interaction.

The limiting behaviour of the solution follows directly from the cumulative profile. Near the origin,
\begin{equation*}
u_{\rm LQG}(r)=\frac{e}{\Lambda}-e(1-q)\pi r^2+\mathcal O(r^4) \ ,
\end{equation*}
which gives
\begin{equation*}
f_{\rm LQG}(r)=1-\frac{2G_NMe}{\Lambda r}+2\pi e(1-q)G_NMr+\mathcal O(r^3) \ .
\end{equation*}
Thus, the leading behaviour is of Schwarzschild type, although the coefficient of the $1/r$ term is rescaled as $M\rightarrow Me/\Lambda$, while the entropy deformation introduces subleading corrections through an exponentially localized dressing of the source. For $(1-q)\Lambda>0$, the cumulative profile decays exponentially at large distances,
\begin{equation*}
u_{\rm LQG}(r) \sim \frac{e}{\Lambda}\exp\!\left[-(1-q)\Lambda\pi r^2\right],\qquad r\rightarrow\infty,
\end{equation*}
implying
\begin{equation*}
m_{\rm LQG}(r)\rightarrow0,
\qquad
G_{\rm LQG}(r)\rightarrow0,
\qquad
f_{\rm LQG}(r)\rightarrow1.
\end{equation*}
Hence, the effective gravitational interaction is exponentially screened, and the spacetime approaches the Minkowski limit asymptotically. The positive distributional source at the origin is exactly compensated by the negative Gaussian dressing cloud, yielding an effective source with vanishing total asymptotic mass.


\section{Conclusions and Discussion}\label{sec:conclusions}

In this work, we have established a general correspondence between generalized black-hole entropy, spacetime geometry, and effective nonlocal gravity for static, spherically symmetric spacetimes. Starting from an arbitrary entropy function, we reconstructed the associated metric through the entropy-geometry relation and showed that the resulting geometry uniquely determines an effective nonlocal operator within the static spherical sector. This construction provides a direct map from generalized entropy to an effective gravitational description without introducing additional fundamental matter fields.

The central ingredient of the formalism is the cumulative mass profile, which simultaneously determines the effective matter distribution, the nonlocal source kernel, and a scale-dependent Newton coupling. Consequently, the entropy-generated anisotropic fluid admits an alternative interpretation as a gravitational polarization cloud produced by the action of an entropy-dependent nonlocal operator on an initially localized source. Within this framework, the effective matter sector and the nonlocal modification of gravity are simply two equivalent representations of the same underlying geometry.

The correspondence also provides a unified framework for comparing different generalized entropy proposals. We demonstrated that each entropy deformation generates a distinct class of nonlocal operator and, consequently, a different gravitational response. In particular, Rényi entropy gives rise to an infrared inverse-Laplacian operator, Tsallis-Cirto and Barrow entropies naturally generate fractional Laplace operators whose orders are determined by their respective deformation parameters, Kaniadakis entropy produces a nonpolynomial exponentially screened kernel, logarithmically corrected entropy replaces the point source with a smooth extended distribution, and the exponentially corrected and LQG-inspired entropies retain a renormalized point source dressed by an exponentially localized cloud. These examples illustrate that generalized entropy corrections should not be viewed as producing a universal ultraviolet modification of gravity, but rather correspond to qualitatively different classes of nonlocal gravitational dynamics.

An immediate consequence of the present construction is the interpretation of the cumulative profile as a running Newton coupling. The associated radial flow is directly related to the effective energy density, providing a simple geometrical measure of gravitational running induced by entropy deformations. In this picture, the Bekenstein-Hawking entropy corresponds to a constant Newton coupling and the identity operator, whereas generalized entropy functions describe nontrivial scale-dependent gravitational interactions.

We have shown that the generalized entropy satisfies $\dd S=\dd A/4G_S(r_+)$ exactly, for any entropy function, so that it is precisely the area law evaluated with the running coupling it itself generates. This resolves the question of whether the reconstructed geometry carries the entropy used to build it: it does, in the scenario in which the gravitational sector is modified, whereas in the scenario in which the geometry is sourced by matter within unmodified Einstein gravity the horizon entropy would be $A/4G_N$. The latter is moreover excluded outright for the R\'enyi and Tsallis-Cirto ($\delta<1$) entropies, which lie below the area law and would therefore require a negative matter entropy. The thermodynamics thus breaks a degeneracy that the field equations alone leave unresolved. We have also identified the conditions under which the reconstruction describes a black hole, namely $S'>0$ and $S''>0$, the latter being equivalent to $T_H>0$; its failure locates degenerate horizons and hence bounds on the mass, an upper bound for the R\'enyi entropy and a lower bound for the logarithmically corrected one, and excludes the Tsallis-Cirto branch with $\delta<1/2$ altogether.\\

The selection of the modified-gravity reading also renders the effective sources internally coherent. Several of the models analysed here, namely the Barrow, Kaniadakis and LQG entropies as well as the Tsallis-Cirto entropy with $\delta>1$, generate a negative effective density for $r>0$, and it is precisely this negative dressing that compensates the distributional contribution at the origin and screens the asymptotic mass. Interpreted as literal matter, such distributions would violate the standard energy conditions while simultaneously being required to carry the entropy bookkeeping of Eq.~\eqref{eq:GSL_decomposition}. Interpreted instead as gravitational polarization induced by a scale-dependent coupling, they carry no such obligation, and their sign merely reflects the direction in which the coupling runs. The reading selected by the thermodynamic argument is therefore the same one that renders the reconstructed sources physically unproblematic.

Although the entropy-geometry correspondence is exact within the static, spherically symmetric sector, it should not be interpreted as a proof that the reconstructed operator defines a unique fundamental theory of nonlocal gravity. The effective source determines only the scalar component of the operator relevant for static point sources, and different generally covariant tensorial actions may reduce to the same spherical kernel. Furthermore, the reduced operator equation employed throughout this work should be regarded as the symmetry-reduced representation of the theory rather than the unrestricted metric variation of a generic nonlocal action. A complete covariant formulation must additionally account for the variation of nonlocal operators, tensorial consistency conditions, and the causal prescription adopted in Lorentzian spacetime.

From a broader perspective, the present results suggest that generalized black-hole entropy may encode information not only about horizon thermodynamics but also about the underlying structure of the gravitational interaction itself. The entropy function determines the effective matter sector, the associated nonlocal operator, the running gravitational coupling, and the resulting spacetime geometry within a single unified framework. This establishes a direct bridge between generalized black-hole thermodynamics and effective nonlocal gravity.

Several important questions remain open. It would be desirable to reconstruct the full covariant nonlocal action corresponding to a given entropy function, investigate the associated spin-2 propagator and ghost structure, analyze causal properties and stability, and extend the formalism beyond static spherical symmetry to rotating black holes, dynamical spacetimes, and cosmological backgrounds. Exploring these directions may provide further insight into the role of generalized entropy in quantum gravity and clarify whether entropy-generated geometries can be understood as effective manifestations of an underlying nonlocal theory of gravitation.


\bibliographystyle{ref}
\bibliography{ref}

\end{document}